\documentclass[aps,pra,twocolumn]{revtex4-2}

\usepackage{dcolumn}
\newcolumntype{w}[1]{D{.}{.}{#1}}
\usepackage{times}
\usepackage{amsmath}
\newcommand{\ddD}[1]{\mathrm{d}^D #1}
\allowdisplaybreaks

\begin{document}
\preprint{Version 2.0}

\title{Recoil corrections to $\mu$H hyperfine splitting}

\author{Andrzej Maro\'n, Mateusz Pa\'ntak,  and Krzysztof Pachucki}

\affiliation{Faculty of Physics, University of Warsaw,
             Pasteura 5, 02-093 Warsaw, Poland}

\date{\today}
\begin{abstract}
This work attempts to present a complete theory of the $\mu$H hyperfine splitting, including all contributions above 1 ppm. 
Quantum electrodynamic and  recoil corrections are calculated directly, 
while the proton structure correction is obtained with the help of the H hyperfine splitting.  
The resulting theoretical prediction for the ground state of $\mu$H is $E_\mathrm{hfs} = 182\,626(5)$ $\mu$eV. 
\end{abstract}

\maketitle

\section{Introduction}
The ground state hyperfine splitting (HFS) in regular hydrogen  has long served as a low-energy test of the Standard Model
of fundamental interactions \cite{Bodwin:88, volotka, SGK:05}. Despite the apparent simplicity of the hydrogen atom, 
the high-precision calculation of two-body effects and the estimation of proton structure corrections remain a challenge. 
In fact, we observe a $2\sigma$ discrepancy between the most recent theoretical predictions \cite{Hevler:26}
and extremely accurate HFS measurements \cite{SGK:00, Hmaser1, Hmaser2}.   
This recent  work \cite{Hevler:26}  improved theoretical predictions by recalculation  of the leading relativistic recoil correction, 
and the not well-known proton structure is most probably the source of this remaining $2\sigma$ discrepancy.
At present, there is no straightforward way to improve the theoretical estimates for the Zemach radius and the proton polarizability. 
Lattice QCD is not yet  able to predict proton properties at the 1\% level.
The only viable approach to improving the hydrogen test of the Standard Model is through the measurement of HFS in another hydrogenic system, 
namely in $\mu$H, where the electron is replaced by a muon \cite{antognini, vacchi}.

In this work, we study QED contributions to the $\mu$H HFS, with emphasis on  nuclear recoil corrections.
These corrections become highly significant here due to the  muon-proton mass ratio being approximately 0.1, 
which is much larger than the 0.0005 ratio in regular hydrogen. 
We aim to identify all contributions larger than 1ppm. Most of them are subsequently calculated, while 
a few remaining ones are only estimated and  left for future investigation. Our work is probably the first attempt
of a comprehensive calculation of $\mu$H HFS. 

\section{Leading order HFS}
Let us introduce the notation before proceeding to the calculations.
The spin-averaged expectation value of an operator $Q$ is denoted by $\langle Q \rangle$. The expectation value
involving the nuclear spin $\vec I$ depends on the total angular momentum $F$, so we denote it by $\langle Q \rangle_F$.
Finally,  $\langle Q\rangle_\mathrm{hfs}$  denotes the difference 
\begin{align}
\langle Q \rangle_\mathrm{hfs} =&\ \langle Q \rangle_{J+1/2} - \langle Q \rangle_{J-1/2}\,. \label{01}
\end{align}
If $Q$ does not involve the nuclear spin, then $\langle Q \rangle_\mathrm{hfs} = 0$.
Therefore, we consistently drop the subscript ``hfs" in $\langle Q \rangle_\mathrm{hfs}$ for operators $Q$ that involve the nuclear spin.

The nuclear magnetic moment
\begin{align}
\vec\mu_I = \frac{q}{2\,M}\,g\,\vec I\,, \label{02}
\end{align}
where $q=-Z\,e$ and $e$ is the electron charge,
can be expressed in terms of the nuclear $g$ factor and the
magnetic moment anomaly $\kappa$ with $g = 2\,(1+\kappa)$.
In a nonrelativistic framework, the interaction between the pointlike nuclear magnetic moment
and that of the muon  leads to a hyperfine splitting of the atomic energy levels given by the expectation value
of the magnetic interaction
\begin{align}
V_\mathrm{hfs} =&\ -\frac{2}{3}\,\vec \mu_I\cdot\vec\mu_\mu\,\delta^{3}(r)
\nonumber \\ =&\
\frac{8}{3}\,\frac{Z\,\alpha}{m_\mu\,M}\,
(1+\kappa)\,(1+a_\mu)\,\vec I\cdot\vec s\,\pi\,\delta^3(r)\,, \label{03}
\end{align}
 evaluated with the nonrelativistic wave function $\phi$
\begin{align}
E_\mathrm{hfs} =&\ \langle \phi| V_\mathrm{hfs}| \phi\rangle
\nonumber \\ =&\
\frac{8}{3}\,\frac{(Z\,\alpha)^4}{n^3}\,\frac{\mu^3}{m_\mu\,M}\,(1+\kappa)\,(1+a_\mu)\,\langle \vec I\cdot\vec s\rangle
\nonumber \\ \equiv&\ E_F\,(1+a_\mu)\,, \label{04}
\end{align}
where  $\mu$ is the reduced mass
\begin{align}
\frac{1}{\mu} =&\ \frac{1}{m_\mu} + \frac{1}{M}\,. \label{05}
\end{align}
Below we investigate all corrections to HFS and express them in terms of $\delta$ defined by $E_\mathrm{hfs} = E_F(1 +\delta)$.

\section{EVP in $\mu$H for a point nucleus}
For a pointlike proton, the largest QED correction  arises from electron vacuum polarization (EVP). 
This is because the spatial size of EVP is on the order of the (muonic) Bohr radius.
In general, vacuum polarization modifies the photon propagator as follows \cite{Itzykson}
\begin{align}
\frac{1}{k^2} \rightarrow \frac{1}{k^2\,\big(1+ \bar\omega(k^2/m_e^2)\big)} 
\approx -\frac{\bar\omega(k^2/m_e^2)}{k^2}\,, \label{06}
\end{align}
where $k^2 = \omega^2-\vec k^{\,2}$. The Coulomb interaction is modified accordingly \cite{rmp}:
\begin{align}
\bigg[\frac{1}{r}\bigg]_\mathrm{vp} =&\ \int \frac{d^3k}{(2\,\pi)^3}\,\frac{4\,\pi}{\vec k^{\,2}}\,[-\bar\omega(-\vec k^{\,2}/m_e^2)]\,e^{i\,\vec k\cdot\vec r}\,. \label{07}
\end{align}
Using  the integral form
\begin{align}
\bar{\omega}(k^2) = &\ 
\frac{\alpha}{\pi}\,k^2 \int_4^\infty\,d(q^2)\frac{1}{q^2\,(q^2-k^2)}\,u(q^2)\,, \label{08}
\end{align}
one obtains
\begin{align}
\bigg[\frac{1}{r}\bigg]_\mathrm{vp} =&\ 
\frac{\alpha}{\pi}\!\int_4^\infty\!d(q^2)\,\frac{u(q^2)}{q^2}\!
\int \frac{d^3k}{(2\,\pi)^3}\,\frac{4\,\pi\,e^{i\,\vec k\cdot\vec r}}{(k^2 + m_e^2\,q^2)}
\nonumber \\ =&\
\frac{\alpha}{\pi}\!\int_4^\infty\!d(q^2)\,\frac{u(q^2)}{q^2}\,\frac{e^{-m_e q\,r}}{r}\,, \label{09}
\end{align}
which is a convenient representation of the EVP potential.
Thus, the calculation of the EVP corrections to the HFS is performed in two steps.
In the first step,  matrix elements with the Coulomb potential $1/r$ 
replaced by a Yukawa potential $\exp(-\rho\,r)/r$ are obtained analytically and denoted by $E(\rho)$.
In the second step, one  numerically evaluates the integral
\begin{align}
E =&\ \frac{\alpha}{\pi}\!\int_4^\infty\!d(q^2)\,\frac{u(q^2)}{q^2}\,E(m_e\,q)\,, \label{10}
\end{align}
where 
\begin{align}
u(q^2) = &\ \frac{1}{3}\sqrt{1-\frac{4}{q^2}}\,\left(1+\frac{2}{q^2}\right)\,. \label{11}
\end{align}

\subsection{$\delta^{(1)}_\mathrm{evp}$}
Let us begin the calculation with the one-loop EVP.
For a massive photon, the spin-spin interaction takes the form
\begin{align}
V_\mathrm{hfs}(\rho) =&\ \frac{8}{3}\,\frac{Z\,\alpha}{m_\mu\,m_p}\,(1+\kappa)\,(1+a_\mu)\,\vec I\cdot\vec s
\nonumber \\ &\times
\left( \pi\,\delta^3(r)-\frac{1}{4}\,\frac{e^{-\rho\,r}}{r^3}\,(\rho\,r)^2 \label{12}
\right)\,.
\end{align}
The correction to the HFS can be expressed as
\begin{align}
&\delta E_\mathrm{hfs}(\rho)  =\    \langle V_\mathrm{hfs}(\rho)\rangle + 2\,\biggl\langle V_\mathrm{hfs}\,
\frac{1}{(E-H)'}\,(-Z\,\alpha)\,\frac{e^{-\rho\,r}}{r}\bigg\rangle
\nonumber \\ &=\ E_F\,(1+a_\mu) \left[ \frac{8}{\kappa+2} 
+ \frac{8 + 8\ln\left(1+\kappa/2\right)}{(\kappa+2)^2} - \frac{16}{(\kappa+2)^3} \right] , \label{13}
\end{align}
where $\kappa = \rho / (\mu\,Z\alpha)$.
The result of numerical integration  
\begin{align}
\delta E_\mathrm{hfs} =&\  \frac{\alpha}{\pi}\,\int_4^\infty \frac{d(\rho^2)}{\rho^2}
\,u\left(\frac{\rho^2}{m_e^2}\right)\,\delta E_\mathrm{hfs}(\rho)
\nonumber \\ =&\ 
E_F\,(1+a_\mu)\, \delta^{(1)}_\mathrm{evp} \label{14}
\end{align}
for the 1S state is
\begin{align}
\delta^{(1)}_\mathrm{evp} =&\ 0.006\,075\,29\,. \label{15}
\end{align}

\subsection{$\delta^{(2)}_\mathrm{evp}$}
$\delta^{(2)}_\mathrm{evp}$ represents the corresponding two-loop EVP corrections, which we obtain using the {\small  PBARSPECTR} code \cite{pbarspectr}.
Namely, we numerically solve the Schr\"odinger equation for a pointlike nucleus with one- and two-loop EVP
\begin{align}
\delta^{(1)}_\mathrm{evp,npert} =&\ 0.006\,089\,78\,, \label{16}
\\
\delta^{(2)}_\mathrm{evp,npert}  =&\ 0.000\,046\,76\,, \label{17}
\end{align}
and we extract $\delta^{(2)}_\mathrm{evp}$ by subtracting $\delta^{(1)}_\mathrm{evp}$
\begin{align}
\delta^{(2)}_\mathrm{evp} =&\ \delta^{(1)}_\mathrm{evp,npert} + \delta^{(2)}_\mathrm{evp,npert} - \delta^{(1)}_\mathrm{evp}
\nonumber \\ =&\ 0.000\,061\,25\,. \label{18}
\end{align}
This partially includes three-loop corrections, but they are expected to be negligibly small.

\subsection{$\delta^{(3)}_\mathrm{rel,evp}$}
This is the one-loop EVP calculated using the Dirac wave function, or more precisely, the relativistic correction to $\delta^{(1)}_\mathrm{evp}$.
The relativistic form of the hyperfine interaction 
\begin{align}
V_\mathrm{hfs}(\vec r) =&\ \frac{e}{4\,\pi}\,\vec \mu_I\cdot\vec\alpha\times\frac{\vec r}{r^3} \label{19}
\end{align}
for a massive photon is
\begin{align}
    V_{\text{vp}}(\vec{r}, \rho) =&\ \frac{e}{4\pi} \vec{\mu} \cdot \left( \vec{\alpha} \times \frac{\vec{r}}{r^3} \right) e^{-\rho r} (1 + \rho r)\,. \label{20}
\end{align}
$\delta E_\mathrm{hfs}(\rho)$ takes an expectation value form similar to that of the nonrelativistic case
\begin{align}
\delta E_\mathrm{hfs}(\rho)  =&\   
\langle V_\mathrm{hfs}(\rho)\rangle + 2\,\biggl\langle V_\mathrm{hfs}\,\frac{1}{(E-H)'}\,(-Z\,\alpha)\,\frac{e^{-\rho\,r}}{r}\bigg\rangle
\nonumber \\ =&\ \delta E_\mathrm{hfs1}(\rho) + \delta E_\mathrm{hfs2}(\rho) \label{21}
\end{align}
with the relativistic wave function $\psi$, the energy $E$, and the Hamiltonian $H$.
The unperturbed ground state wave function $\psi$ is described by the spherical Dirac spinor,
\begin{align}
    \psi(\vec{r}) &= \frac{1}{\sqrt{4\pi}} \begin{pmatrix} g(r) \chi_m \\ -i f(r) (\vec{\sigma} \cdot \hat{r}) \chi_m \end{pmatrix}\,, \label{22} \\
    g(r) &= N r^{\gamma-1} e^{-Z\alpha r}\,, \quad f(r) = -\sqrt{\frac{1-\gamma}{1+\gamma}} \, g(r)\,, \label{23}
\end{align}
where we adopt atomic units ($m=1$) and $\gamma = \sqrt{1-(Z\alpha)^2}$.
The first part $\delta E_\mathrm{hfs1}(\rho)$ is
\begin{align}
 \delta E_\mathrm{hfs1}(\rho) =&\   \langle \psi | V_\mathrm{hfs}(\rho) | \psi \rangle = E_{F\infty}\,\delta^{(1+)}_\mathrm{evp1}\,, \label{24}
 \end{align}
where
\begin{equation}
    E_{F\infty} = \frac{8}{3} (Z\alpha)^4 \frac{m^2}{M} \frac{g}{2} \langle \vec{I} \cdot \vec{s} \rangle \label{25}
\end{equation}
is the nonrecoil limit of the Fermi splitting. Using Eqs. (\ref{22}) and (\ref{23}), one obtains
\begin{align}
\delta^{(1+)}_\mathrm{evp1}(\rho)  =&\  \frac{1}{2(Z\alpha)^3} \int_0^\infty dr \, (-2)g(r) f(r)\, e^{-\rho r} \big( 1 + \rho r \big)
\nonumber \\ =&\
\frac{1}{\gamma(2\gamma-1)} \left( \frac{2}{\tau+2} \right)^{2\gamma} (1+\gamma\,\tau)\,, \label{26}
\end{align}
where $\tau = \rho / (m_\mu\,Z\alpha)$.
Considering the second part, $ \delta E_\mathrm{hfs2}(\rho)$, 
we note that within the subspace of the Dirac quantum number  $\kappa=-1$ ,  $V_\mathrm{hfs}$ can be simplified to
\begin{align}
\langle V_\mathrm{hfs}\rangle
=&\ \frac{2}{3}\frac{e}{4\,\pi}\,  \langle\vec \mu_I\cdot\vec\sigma\rangle_\mathrm{hfs}\; \bigg\langle i\,\frac{\vec\gamma\cdot\vec r}{r^3}\bigg\rangle\,. \label{27}
\end{align}
Therefore,  $ \delta E_\mathrm{hfs2}(\rho)$ can be written as
\begin{align}
\delta E_\mathrm{hfs2}(\rho) = &\ 2 \bigg\langle \psi \bigg| -\frac{Z\alpha}{r} e^{-\rho r} 
\bigg| \delta \psi \bigg\rangle\,\frac{2}{3}\,\frac{e}{4\,\pi}\,\langle \vec \mu_I\cdot\vec\sigma\rangle_\mathrm{hfs}
\nonumber \\ =&\
E_{F\infty}\,\delta^{(1+)}_\mathrm{evp2}\,, \label{28}
\end{align}
where
\begin{align}
\delta\psi(\vec r) =&\ \frac{1}{(E-H)'}\, i\,\frac{\vec\gamma\cdot\vec r}{r^3}\,\psi(\vec r)
 = \begin{pmatrix}W(r) \chi_m \\ -iZ(r) (\vec{\sigma} \cdot \hat{r}) \chi_m \end{pmatrix}. \label{29}
\end{align}
Using the Shabaev method \cite{shabaev1, shabaev2}, one obtains
\begin{widetext}
\begin{align}
    W(r) &= C \left[ \frac{2(Z \alpha)^3}{\gamma} g - 3(1+\gamma) f - \frac{1+2\gamma}{r} g - 2Z\alpha(2\gamma+1) \left( \frac{\Psi(2\gamma+1)}{\gamma} + \gamma + 1 - Z\alpha r - \frac{\ln(2 Z \alpha r)}{\gamma} - \frac{1}{2\gamma} \right) g \right], \label{30}\\
    Z(r) &= C \left[\frac{2(Z \alpha)^3}{\gamma} f - 3(1-\gamma) g - \frac{3(1+2\gamma)}{r} f - 2Z\alpha(2\gamma+1) \left( \frac{\Psi(2\gamma+1)}{\gamma} + \gamma + 1 - Z\alpha r - \frac{\ln(2 Z \alpha r)}{\gamma} + \frac{1}{2\gamma} \right) f \right], \label{31}
\end{align}
where $C = 1/[4(Z\alpha)^2 - 3]$. The corresponding dimensionless contribution is 
\begin{align}
\delta^{(1+)}_\mathrm{evp2}  =&\  \frac{1}{(Z\alpha)^2} \int_0^\infty dr \, \big[ W(r)g(r) + Z(r)f(r) \big] r e^{-\rho r}
\nonumber \\ =&\  
\frac{\beta^{-2\gamma}}{\gamma(2\gamma-1)} \Bigg[ \frac{2(2-\gamma)}{2\gamma-1} \beta 
+ \frac{2\gamma^3 + 2\gamma^2 - 2\gamma + 1}{\gamma^2} + \frac{2}{\gamma}\ln\beta - \frac{2\gamma}{\beta} \Bigg]\, ,  \label{32}
\end{align}
where $\beta = 1+\tau/2$. The sum of both parts $\delta^{(1+)}_\mathrm{evp1} + \delta^{(1+)}_\mathrm{evp2} = \delta^{(1)}_\text{evp}  + \delta^{(3)}_\text{rel,evp}  +\ldots$
after expansion in $Z\,\alpha$ is
\begin{align}
\delta^{(3)}_{\text{rel,evp}} =&\frac{\alpha}{\pi}\,(Z\alpha)^2\,\int_4^\infty \frac{d(\rho^2)}{\rho^2}
\,u\left(\frac{\rho^2}{m_e^2}\right)  \Bigg[ \frac{8 + 4\ln\beta}{\beta} + \frac{3 + 6\ln\beta + 2\ln^2\beta}{\beta^2} - \frac{2 + 2\ln\beta}{\beta^3} \Bigg]\,
\nonumber \\ =&\
 1.15 \text{ ppm}\,. \label{33}
\end{align}
In Sec. V, we calculate the EVP combined with nuclear recoil or finite nuclear size effects, and in Sec. VII we consider it combined with $\mu$VP and $\mu$SE,
but before this we must briefly describe an approach for nuclear recoil corrections. 

\section{Two-photon-exchange forward scattering amplitude}
To calculate corrections beyond the pointlike, static nucleus approximation, we first
consider the two-photon-exchange correction to the HFS.
We closely follow our previous work in Ref. \cite{pachucki:22} and use the temporal gauge 
\begin{align}
E^{(5)}_{\rm hfs} =&\ \frac{i}{2}\,\,\phi^2(0)
\int\frac{d^4k}{(2\,\pi)^4}\,\frac{1}{k^4}\,
\biggl(\delta^{ik}-\frac{k^i\,k^k}{\omega^2}\biggr)\,
\biggl(\delta^{jl}-\frac{k^j\,k^l}{\omega^2}\biggr)\,t^{ji}\,T^{kl}\,. \label{34}
\end{align}
For a pointlike spin-$1/2$ particle
\begin{align}
t^{ji} =&\ e^2\biggl[
\langle\bar u|\gamma^j\frac{1}{\not\!t\;- \not\!k-m}\,\gamma^i|u\rangle +
\langle\bar u|\gamma^i\frac{1}{\not\!t\;+ \not\!k-m}\,\gamma^j|u \rangle \biggr]\,, \label{35}
\end{align}
and for a finite size  spin-$1/2$ particle
\begin{align}
T^{kl} =&\ (Z\,e)^2\biggl[
\langle\bar u|\Gamma^k(k)\frac{1}{\not\!t\;- \not\!k-m}\,\Gamma^l(-k)|u \rangle +
\langle\bar u |\Gamma^l(-k)\frac{1}{\not\!t\;+ \not\!k-m}\,\Gamma^k(k)|u\rangle\biggr]\,, \label{36}
\end{align}
\end{widetext}
where $t$ is the four-momentum at rest, $t=(m,\vec 0)$, and
\begin{align}
\Gamma^\mu(k) =&\ \gamma^\mu\,F_1 + \frac{i}{2\,M}\,\sigma^{\mu\nu}\,k_\nu\,F_2 \,. \label{37}
\end{align}
Using the decomposition in terms of scalar functions $t_i$ and $T_i$ with $i=1,2$
\begin{align}
t^{ji} =&\ i\,\epsilon^{ijk}\,e^2\,\omega\,\bigg(t_1\,s^k +  t_2\,\frac{k^k}{\vec k^{\,2}}\,\vec k\cdot \vec s\bigg)\,, \label{38}
 \\ 
T^{ji} = &\ i\,\epsilon^{ijk} (Ze)^2 \omega \left( T_1 I^k + T_2 \frac{k^k}{\vec k^{\,2}} \vec{k} \cdot \vec{I} \right)\,, \label{39}
\end{align}
one obtains for the lepton
\begin{align}
t_1 =&\ \frac{4\,k^2}{(k^2-2\,m\,\omega)\,(k^2+2\,m\,\omega)}\,,\label{40} \\
t_2 =&\ 0\,, \label{41}
\end{align}
and for the nucleus
\begin{align}
T_1(-k^2,\omega) =&\ \frac{4 F_1\,M^2\,[F_1\,k^2 + F_2\,(k^2 + \omega^2)] - F_2^2\,k^4}
{(k^2 - 2 M \omega)(k^2 + 2 M \omega)M^2}\,, \label{42}\\
T_2(-k^2,\omega) =&\ \frac{4\,(k^2-\omega^2)\, F_2 (F_1 + F_2)}{(k^2 - 2 M \omega)(k^2 + 2 M \omega)}\,. \label{43}
\end{align}
The two-photon exchange correction takes the form
\begin{align}
E^{(5)}_{\rm hfs}  =&\
i\,\phi^2(0)\,(4\,\pi\,Z\,\alpha)^2\,\frac{\vec I\cdot\vec s}{3}\,\int\frac{d^4k}{(2\,\pi)^4}
\nonumber \\ &\times
\biggl[\frac{2}{k^2}\,t_1\,T_1 + \frac{\omega^2}{k^4}\,(t_1+t_2)\,(T_1+T_2)\biggr]\,. \label{44}
\end{align}
Since the form factors are functions of $-k^2$, one performs a Wick rotation $\omega = i\,k_0$, and $k^2\rightarrow -k^2$,
and averages over the three-dimensional sphere in Euclidean space
\begin{align}
A[f] \equiv&\ \int \frac{d\,\Omega_k}{2\,\pi^2}\,f(k,k_0) 
\nonumber \\ =&\
\frac{2}{\pi}\int_0^\pi d\phi\,(\sin\phi)^2\, f\big(k,k\,\cos\phi\big)\,. \label{45}
\end{align}
For instance,
\begin{align}
A\biggl[\frac{1}{k^4+4\,M^2\,k_0^2}\biggr] =&\  \frac{2}{k^4}\,\frac{1}{1 + \sqrt{1 + 4\,M^2/k^2}}\,. \label{46}
\end{align}
We thus obtain
\begin{align}
E^{(5)}_{\rm hfs}  =&\
\frac{16}{3}\,\phi^2(0)\,(Z\,\alpha)^2\,\frac{\vec I\cdot\vec s}{3}\,\int\frac{d^4k}{(2\,\pi)^4}\,
\nonumber \\ &\times
A\biggl[\frac{2}{k^2}\,t_1\,T_1 + \frac{k_0^2}{k^4}\,(t_1+t_2)\,(T_1+T_2)\biggr]\,. \label{47}
\end{align}
The low $k$ asymptotic behavior of the integrand 
\begin{align}
A[t_\mathrm{1}\,T_1] \approx &\ \frac{16\,(1+\kappa)}{M+m}\,\frac{1}{k^3} \label{48}
\end{align}
should be subtracted out \cite{iddings:65, pachucki:96}, yielding 
\begin{widetext}
\begin{align}
    E^{(5)}_{\text{hfs}} = &\ - \frac{16}{3 }\,(Z\,\alpha)^2 \frac{\phi^2(0)}{M\,m} \, \vec I \cdot\vec s \int \frac{dk}{k} \, 
    \bigg[ -4\,\frac{m}{k}\,G_E(k^2)G_M(k^2)  + 4\,(1+\kappa)\,\frac{m}{k}  + T_{\text{rec}}(k) - 4\,(1+\kappa)\,\frac{m}{k} \frac{m}{M+m}\bigg],
    \label{49}
\end{align}
where  the recoil part $T_{\text{rec}}$ is
\begin{align}
T_{\text{rec}}(k) =&\ \frac{4\,m}{k} G_E(k^2) G_M(k^2) + \frac{m}{M}\,\left[ \frac{k^2}{8\,m^2} - 2\,\left( 1 + \sqrt{1 + \frac{4m^2}{k^2}} \right)^{-1} \left( \frac{k^2}{8\,m^2} - 1 \right) \right] F_2^2(k^2) \nonumber \\ &\ 
+ \frac{M\,m}{M^2 - m^2}\,\left( 1 + \sqrt{1 + \frac{4\,M^2}{k^2}} \right)^{-1}\, \left[ \left( 1 - \frac{8M^2}{k^2} \right) F_1(k^2) + 3 F_2(k^2) \right]\,G_M(k^2) 
\nonumber \\ &\
- \frac{M\,m}{M^2 - m^2}\,\left( 1 + \sqrt{1 + \frac{4\,m^2}{k^2}} \right)^{-1}\,  \left[ \left( 1 - \frac{8m^2}{k^2} \right) F_1(k^2) + 3 F_2(k^2) \right]\,G_M(k^2)\,. \label{50}
\end{align}
\end{widetext}
The Sachs electric $G_E$ and magnetic $G_M$ form factors are related to $F_1$ and $F_2$ by
\begin{align}
G_E(k^2) =&\ F_1(k^2) - \frac{k^2}{4\,M^2}\,F_2(k^2)\,, \label{51}\\
G_M(k^2) =&\ F_1(k^2)+F_2(k^2)\,, \label{52}
\end{align}
with the normalization $G_M(0) = 1+\kappa = g/2$.
For a pointlike nucleus  $F_1=1$ and $F_2=0$, we obtain
\begin{align}
E^{(5)}_{\text{point,hfs}}=-8 \,(Z\, \alpha)^2 \frac{\phi^2(0)}{M^2 - m^2} \, (\vec{I} \cdot \vec{s})\, \ln \frac{M}{m}, \label{53}
\end{align}
in agreement with a well-known result \cite{eides:01} for the recoil correction to the hyperfine splitting.
For a finite size nucleus in the nonrecoil limit,  we obtain the Zemach correction \cite{Zemach}
\begin{align}
\delta^{(1)}_\mathrm{fns}=&\ \frac{2\,Z\,\alpha\,m}{\pi^2}\,
\int\frac{d^3k}{k^4}\,\biggl[\frac{G_E(k^2)\,G_M(k^2)}{1+\kappa}-1\biggr]
\nonumber \\ =&\  
-2\,Z\,\alpha\, m\,r_{\rm Z}\,.
\label{54}
\end{align}
It is convenient to express this finite nuclear size correction in terms of the Zemach radius $r_{\rm Z}$ \cite{Zemach}, defined as 
\begin{align}
r_{\rm Z} = \int d^3 r_1 \int d^3 r_2\,\rho_E(r_1)\,\rho_M(r_2)\,|\vec r_1-\vec r_2|, \label{55}
\end{align}
where $\rho_E$ and $\rho_M$ are the Fourier transforms of $G_E$ and $G_M/(1+\kappa)$, respectively.
Using the dipole parametrization for the nuclear form factors $\rho_E = \rho_M = \rho$ with
\begin{align}
\rho(k^2) =&\ \frac{\Lambda^4}{(\Lambda^2+k^2)^2}\,, \label{56}
\end{align}
one finds $r_Z = 35/(8\,\Lambda)$.
Using the dipole parametrization with $\Lambda$ adjusted to the Zemach radius $r_Z = 1.054 \, \text{fm}$, 
we obtain a recoil correction of $\delta^{(1)}_\mathrm{rec} = 0.001\,667$. This is not very different from the more accurate value of $\delta^{(1)}_\mathrm{rec} =0.001\,672(3)$ 
obtained using more realistic proton form factors \cite{AA2022a}. Therefore, our subsequent calculations of vacuum polarization(VP) and 
self-energy (SE) corrections to the two-photon exchange amplitude
employ the dipole approximation for the proton form factors.

\section{VP combined with recoil and FNS}
The radiative recoil $Z\,\alpha^2\,m/M\,E_F$ correction has previously been studied only for muonium HFS \cite{eides:01}, 
but not for a finite-size nucleus with an arbitrary $g$ factor.
Here we derive formulas without expansion in $m/M$ and $\Lambda/M$, expressing the result in terms of a one-dimensional integral.
The electron vacuum polarization modifies the photon propagator according to Eq. (\ref{06}).
The explicit formula for $\bar\omega$ is given by \cite{Itzykson}
\begin{align}
\bar\omega(-k^2) =&\  -\frac{\alpha}{3\,\pi}\,\biggl\{\frac{1}{3} + 2\biggl(1-\frac{2}{k^2}\biggr)
\nonumber \\ &\times
\biggl[\sqrt{1+\frac{4}{k^2}}\,\mathrm{arccoth}\sqrt{1+\frac{4}{k^2}}-1\biggr]\biggr\}. \label{57}
\end{align}
The VP correction due to a lepton of mass $m'$ ($m'\neq m_e$ for $\mu$H) 
can be easily obtained from the two-photon exchange amplitude (without subtracting the low-$k$ asymptotic behavior)
\begin{align}
    E^{(6)}_\mathrm{vp} = &\ \frac{16}{3 }\,(Z\,\alpha)^2 \frac{\phi^2(0)}{M\,m} \, \vec I \cdot\vec s \int \frac{dk}{k}
    \nonumber \\ \times&
    \bigg[ 4\,\frac{m}{k}\,G_E(k^2)G_M(k^2) - T_{\text{rec}}(k)\bigg]\,(-2)\,\bar\omega\Big(-\frac{k^2}{m'^2}\Big)
    \nonumber \\
=&\  E^{(6)}_\mathrm{vp,point}   + E^{(6)}_\mathrm{vp,fns} + E^{(6)}_\mathrm{vp,rec}\,. \label{58}
\end{align}
$E^{(6)}_\mathrm{vp,point}$ arises from the low $k$ asymptotic behavior of the integrand
\begin{align}
 E^{(6)}_\mathrm{vp,point} =&\ \frac{16}{3 }\,(Z\,\alpha)^2 \frac{\phi^2(0)}{M\,m} \, \vec I \cdot\vec s \int \frac{dk}{k^2}
    \nonumber \\ &\times 
4\,(1+\kappa)\,\mu\,(-2)\,\bar\omega\Big(-\frac{k^2}{m'^2}\Big)
\nonumber \\ =&\ E_F\,\frac{3}{4}\,Z\,\alpha^2\,\frac{\mu}{m'}, \label{59}
\end{align}
where $\mu$ is the reduced mass. $E^{(6)}_\mathrm{fns,vp}$ is the finite size correction in the nonrecoil limit,
\begin{align}
E^{(6)}_\mathrm{vp,fns} = &\
E_F\,\frac{2\,Z\,\alpha\,m}{\pi^2}\,\int\frac{d^3k}{k^4}\,\biggl[\frac{G_E(k^2)\,G_M(k^2)}{1+\kappa}-1\biggr]
\nonumber \\ &\times
(-2)\,\bar\omega\Big(-\frac{k^2}{m'^2}\Big)
\nonumber \\ \approx&\
-E_F\,2\,Z\,\alpha\,m\, r_Z
 \,\frac{\alpha}{\pi}\,\biggl(\frac{2}{3}\,\ln\frac{\Lambda^2}{m'^2} - \frac{634}{315}\biggr), \label{60}
\end{align}
where the dipole parametrization of nuclear form factors is assumed, in agreement with Ref. \cite{sgk:97}.
Finally,  $E^{(6)}_{\text{rec,vp}}$ is the recoil VP correction
\begin{align}
    E^{(6)}_{\text{vp,rec}} =&\ - \frac{16}{3 }\,(Z\,\alpha)^2 \frac{\phi^2(0)}{mM} \, \vec I \cdot\vec s \int \frac{dk}{k}
\nonumber \\ \times&    
    \bigg[ T_{\text{rec}}(k) - 4\,(1+\kappa)\,\frac{m}{k}\,\frac{m}{M+m}\bigg]\, (-2)\,\bar\omega\Big(-\frac{k^2}{m'^2}\Big). \label{61}
\end{align}
Using Eq. (\ref{58}), we obtain the following for $\mu$VP in H 
\begin{align}
    \delta_\mathrm{\mu vp,point}(\mathrm{H}) =&\ 0.193\, \, \text{ppm}, \label{62} \\
    \delta_\mathrm{\mu vp,fns}(\mathrm{H}) =&\ -0.121\, \, \text{ppm}, \label{63} \\
    \delta_\mathrm{\mu vp,rec}(\mathrm{H}) =&\ -0.001 \, \, \text{ppm}. \label{64}
\end{align}

In the case of EVP in $\mu$H,  $E^{(6)}_\mathrm{vp,point}$ is treated separately, because the scattering approximation is not valid here.
The calculations for a pointlike nucleus are described in Sec. III.
Here we consider only $E^{(6)}_\mathrm{vp,fns}$ and $ E^{(6)}_\mathrm{vp,rec}$. 
Using Eqs. (\ref{59}) and (\ref{61}), respectively, we obtain
\begin{align}
    \delta_{\mathrm{evp,fns}}(\mathrm{\mu H})=&\ -149.81\, \, \text{ppm}\,, \label{65} \\
    \delta_{\mathrm{evp,rec}}(\mathrm{\mu H})=&\ 25.24 \, \, \text{ppm}\,. \label{66}
\end{align}
The final cases to consider  are EVP H and $\mu$VP $\mu$H, where the VP particle is the same as the one in the atom.
The pointlike VP contribution in the nonrecoil limit is already included in $\delta_\mathrm{QED}$, so we rearrange the remaining corrections as follows.
FNS VP in the nonrecoil limit is given by Eq. (\ref{61}) with $m'=m$, 
while the recoil correction is redefined and includes the recoil part from $E^{(6)}_\mathrm{vp,point}$, namely
\begin{align}
    E^{(6)}_{\mathrm{evp,rec}} =&\  - \frac{16}{3 }\,\alpha^2 \frac{\phi^2(0)}{m\, m_p} \, \vec I \cdot\vec s \int \frac{dk}{k} \,  T_{\text{rec}}(k)\, (-2)\,\bar\omega\Big(-\frac{k^2}{m^2}\Big). \label{67}
\end{align}
Our numerical results for H are
\begin{align}
    \delta_{\mathrm{evp,fns}}(\mathrm{H})=&\ -0.725 \, \, \text{ppm},  \label{68}\\
    \delta_{\mathrm{evp,rec}}(\mathrm{H})=&\ -0.032 \, \, \text{ppm}, \label{69}
\end{align}
and for $\mu$H 
\begin{align}
    \delta_{\mathrm{\mu vp, fns}}(\mathrm{\mu H})=&\ -25.10 \, \, \text{ppm}, \label{70} \\
    \delta_{\mathrm{\mu vp, rec}}(\mathrm{\mu H})=&\ -1.18 \, \, \text{ppm}. \label{71}
\end{align}

\section{SE combined with recoil and FNS}
The lepton SE correction is obtained by replacing $t^{ji}$ with the self-energy corrected tensor $t^{ji}_\mathrm{se}$
in Eqs. (\ref{34}), (\ref{38}), and (\ref{44}): 
\begin{widetext}
\begin{align}
E^{(6)}_{\rm se} =&\ \frac{i}{2}\,\phi^2(0)\!
\int\frac{d^4k}{(2\,\pi)^4}\,\frac{1}{k^4}\,
\biggl(\delta^{ik}-\frac{k^i\,k^k}{\omega^2}\biggr)\,
\biggl(\delta^{jl}-\frac{k^j\,k^l}{\omega^2}\biggr)\,
t_\mathrm{se}^{ji}\,T^{kl}\,,
\nonumber \\ =&\
i\,\phi^2(0)\,(4\,\pi\,Z\,\alpha)^2\,\frac{\vec I\cdot\vec s}{3}\,\int\frac{d^4k}{(2\,\pi)^4}\,
\biggl[\frac{2}{k^2}\,t_\mathrm{1se}\,T_1 + \frac{\omega^2}{k^4}\,(t_\mathrm{1se}+t_\mathrm{2se})\,(T_1+T_2)\biggr]\,, \label{72}
\end{align}
where the arguments of the lepton and proton factors are $(-k^2\,,\omega)$.
The self-energy corrected lepton line is (for its calculation see the Appendix)
\begin{align}
t_\mathrm{1se}(-k^2,\omega) = &\ \frac{\alpha}{2\,\pi}\,\biggl\{  
-\frac{4}{k^2 + 2\,\omega} - J\,\bigg[ \frac{1}{2} + \omega + \frac{8\,(1 + \omega)}{k^2 + 2\,\omega} 
+ \frac{\omega\,(1 + 2\,\omega)\,(2 + \omega)}{2\,(k^2 - \omega^2)}\bigg]
\nonumber \\ &\   
- \biggl[1 + \frac{8}{k^2 + 2\,\omega} + \frac{1 + \omega}{2\,(1 + k^2 + 2\,\omega)\,(k^2 - \omega^2)} 
  + \frac{-1 + 3\,\omega + 2\,\omega^2}{2\,(k^2 - \omega^2)}\biggr]\,\ln(-k^2 - 2\,\omega) 
\nonumber \\ &\
+ \biggl[2 + \frac{18}{k^2 + 2\,\omega} + \frac{16}{(k^2-4)\,(k^2 + 2\,\omega)} 
+ \frac{2\,\omega^2}{k^2 - \omega^2}\biggr]\,\arcsin\left(\frac{k}{2}\right)\,\sqrt{\frac{4}{k^2}-1}
+(\omega \rightarrow -\omega)
\biggr\}, \label{73} \\
t_\mathrm{2se}(-k^2,\omega) = &\ \frac{\alpha}{2\,\pi}\,\biggl\{   
- \frac{1 + \omega}{\omega\,(1 + k^2 + 2\,\omega)} + 
  J\,\biggl[ \frac{k^2}{\omega} + \frac{8 + 15\,\omega + 6\,\omega^2}{2\,\omega} + \frac{3\,\omega\,(2 + 5\,\omega + 2\,\omega^2)}{2\,(k^2 - \omega^2)}\biggr]
\nonumber \\ &\  
 + \biggl[\frac{4 + 3\,\omega}{\omega} -\frac{1 + \omega}{\omega\,(1 + k^2 + 2\,\omega)^2}  
    - \frac{(1 + \omega)\,(2 + \omega + 2\,\omega^2)}{2\,\omega\,(1 + k^2 + 2\,\omega)\,(k^2 - \omega^2)} 
    + \frac{2 - \omega + 9\,\omega^2 + 6\,\omega^3}{2\,\omega\,(k^2 - \omega^2)}\biggr]
     \,\ln(-k^2 - 2\,\omega)  
\nonumber \\ &\  
  + \biggl[-6 + \frac{2}{k^2 + 2\,\omega} - \frac{6\,\omega^2}{k^2 - \omega^2} \biggr]\,\arcsin\left(\frac{k}{2}\right)\,\sqrt{\frac{4}{k^2}-1}
 +(\omega \rightarrow -\omega) \biggr\}, \label{74}
\end{align}
and $J$ is a master integral defined in the Appendix. The nuclear factors $T_1$ and $T_2$ are defined in Eqs. (\ref{42}) and (\ref{43}).
After performing a Wick rotation, we average over the three-dimensional sphere. The low $k$ asymptotic behavior of the integrand
\begin{align}
A[t_\mathrm{1se}\,T_1] \approx &\ \frac{16\,a_e\,(1+\kappa)}{M+m}\,\frac{1}{k^3} \label{75}
\end{align}
is subtracted out, because it corresponds to the already included lower order term, thus
\begin{align}
E^{(6)}_{\rm se} =&\ 
\phi^2(0)\,(4\,\pi\,Z\,\alpha)^2\,\frac{\vec I\cdot\vec s}{3}\,\int\frac{d^4k}{(2\,\pi)^4}\,
A\biggl[\frac{2}{k^2}\,t_\mathrm{1se}\,T_1 + \frac{k_0^2}{k^4}\,(t_\mathrm{1se}+t_\mathrm{2se})\,(T_1+T_2) - \frac{32\,a_e\,(1+\kappa)}{M+m}\,\frac{1}{k^5}\biggr]
\nonumber \\ =&
E^{(6)}_{\rm se,point} + E^{(6)}_{\rm se,fns} + E^{(6)}_{\rm se,rec}. \label{76}
\end{align}
The nonrecoil contribution for a pointlike nucleus is
\begin{align}
E^{(6)}_{\rm se,point} =&\ 
-\phi^2(0)\,(4\,\pi\,Z\,\alpha)^2\,\frac{8\,(1+\kappa)}{3\,M}\,\vec I\cdot\vec s\,\int\frac{d^4k}{(2\,\pi)^4}\,\frac{1}{k^3}\,\biggl[t_\mathrm{1se}(k^2,0)  + \frac{4\,a_e}{k^2}\biggr]\,, \label{77}
\end{align}
where
\begin{align}
t_\mathrm{1se}(k^2,0) =&\ \frac{\alpha}{2\,\pi}\,\biggl\{
\frac{8}{k^2} + \bigg(\frac{16}{k^2} - 1\bigg) \,J(k^2,0) 
+ 4\,\bigg( 1 -\frac{5}{k^2} - \frac{28}{k^4}\bigg)\,\frac{\mathrm{arcsinh}\big(\frac{k}{2}\big)}{\sqrt{1 + \frac{4}{k^2}}} 
- \bigg(2-\frac{16}{k^2} + \frac{1}{k^2-1} \bigg)\,\ln(k^2)\biggr\}. \label{78}
\end{align}
\end{widetext}
Numerical integration yields $\delta_\mathrm{se,point} =-136.16 \text{ ppm}$, in agreement with the known \cite{eides:01}
analytic result $\delta_\mathrm{se,point} =Z\alpha^2\,(\ln 2 - \frac{13}{4})$, which is included in $\delta^{(2)}$ in Eq. (\ref{90}).  
The nonrecoil finite nuclear size contribution, using $t_\mathrm{1se} = 5/k^2\,\alpha/\pi + o(k^{-4})$ is
\begin{align}
E^{(6)}_{\rm se,fns} =&\ 
-\phi^2(0)\,(4\,\pi\,Z\,\alpha)^2\,\frac{2\,\vec I\cdot\vec s}{3\,M}\,\int\frac{d^4k}{(2\,\pi)^4}\
\nonumber \\ &\times
\frac{4\, \big(G_E\,G_M -(1+\kappa)\big)}{k^3}\,t_\mathrm{1se}(k^2,0) \label{78p}
 \\ =&\
-2\,Z\,\alpha\,m\,r_Z\,E_F\,\frac{\alpha}{\pi} \Big[-\frac{5}{4}+ O\Big(\frac{m^2}{\Lambda^2}\Big)\Big]\,, \label{79}
\end{align}
in agreement with Ref. \cite{sgk:97}.  

Using Eq. (\ref{78p}),  we obtain for muonic hydrogen
\begin{align}
    \delta_{\mathrm{se,fns}}(\mu\mathrm{H})=16.18 \text{ ppm}, \label{80}
\end{align}
which significantly differs from the result obtained by omitting the $O(m^2/\Lambda^2)$ terms $\delta_{\mathrm{se,fns}}=23.92 \text{ ppm}$.
For hydrogen, we obtain 
\begin{align}
     \delta_{\mathrm{se,fns}}(\mathrm{H})=0.115 \text{ ppm}, \label{81}
\end{align}
which differs very slightly from the result obtained by omitting the $O(m^2/\Lambda^2)$ terms $\delta_{\mathrm{se,fns}}=0.116 \text{ ppm}$.

The nuclear recoil contribution is
\begin{align}
E^{(6)}_{\rm se,rec} =&\ 
\phi^2(0)\,(4\,\pi\,Z\,\alpha)^2\,\frac{\vec I\cdot\vec s}{3}\,\int\frac{d^4k}{(2\,\pi)^4}\,\frac{1}{k^4}
\nonumber \\ &\times
\biggl[2\,k^2\,t_\mathrm{1se}\,T_1 
+ k_0^2\,(t_\mathrm{1se}+t_\mathrm{2se})\,(T_1+T_2) 
\nonumber \\ &\hspace*{-10ex}
+ \frac{8\,k}{M}\,t_\mathrm{1se}(k^2,0)\,G_E(k^2)\,G_M(k^2)
+ \frac{32\,a_e\,(1+\kappa)\,m}{k\,M\,(M+m)} \biggr]. \label{82}
\end{align}
This integral is evaluated numerically, yielding the following results 
\begin{align}
    \delta_{\mathrm{se,rec}}(\mu\text{H} )=16.48 \text{ ppm}\,, \label{83}
\end{align}
\begin{align}
    \delta_{\mathrm{se,rec}}(\text{H} )=0.104 \text{ ppm}\,, \label{84}
\end{align}
where $\Lambda$ has been adjusted to match the Zemach radius $r_\mathrm{Z}$.

\section{EVP combined with $\mu$VP and SE}
The combined electronic and muonic vacuum polarizations contribution in the nonrecoil limit for a pointlike nucleus
is
\begin{align}
    E^{(7)}_\mathrm{\mu vp,evp} = &\
E_F\,\frac{2\,Z\,\alpha\,m_\mu}{\pi^2}\,6\,
\int\frac{d^3k}{k^4}\,\bar\omega\Big(-\frac{k^2}{m_e^2}\Big)\,\bar\omega\Big(-\frac{k^2}{m_{\mu}^2}\Big), \label{85}
\end{align}
from which we obtain
\begin{align}
    &\delta^{(3)}_{\mu \mathrm{vp, evp}} \,(\mu \text{H})=1.17 \text{ ppm}. \label{86}
\end{align}
Including FNS, this correction decreases to 0.32 ppm, which indicates the significance of the FNS effect.
Here we neglect FNS for consistency, as all $\alpha^3$ corrections are calculated for a point nucleus,
and estimate the unknown $\delta^{(3)}_\mathrm{fns} = \pm 2$ ppm, 
while $\delta^{(2)}_\mathrm{fns}$ is calculated separately for the VP, SE, and relativistic parts.

Another correction is the muon one-loop self-energy combined with EVP
inserted into the exchanged photons between the lepton and the nucleus. 
In the nonrecoil limit, this correction is given by
\begin{align}
E^{(7)}_{\rm se,evp1} =&\ 
-\phi^2(0)\,(4\,\pi\,Z\,\alpha)^2\,\frac{2\,(1+\kappa)\,\vec I\cdot\vec s}{3\,M}
\nonumber \\ &\hspace*{-9ex}
\times\int\frac{d^3k}{(2\,\pi)^3}\,\frac{1}{k^2}\,
\Big(t_\mathrm{1se}(k^2,0) + \frac{4\,a_\mu}{k^2}\Big)\, (-2) \, \bar\omega\Big(-\frac{k^2}{m_e^2}\Big). \label{87}
\end{align}
After numerical integration, we obtain
\begin{align}
    &\delta^{(3)}_{\mathrm{se, evp}1} \,(\mu \text{H})=-1.70 \text{ ppm}, \label{88}
\end{align}
and FNS would decrease this correction to $-1.37$ ppm. 

\section{Summary of hyperfine splitting in muonic hydrogen}

\begin{table}
\caption{Contributions to HFS in $\mu$H, constants are from Ref. \cite{codata22}, $g_p = 5.585\,694\,6893(16) $, $\nu_F = 44\,114\,600.4(2.0)$ MHz,  $E_F = 0.182\,443\,32\,8(8)$ eV,
 and $a_\mu$ is the muon magnetic moment anomaly.}
\begin{ruledtabular}
\begin{tabular}{lw{0.11}l}
\multicolumn{1}{l}{Term} &
        \multicolumn{1}{c}{Value} &
            \multicolumn{1}{c}{Reference}
\\
\hline\\[-5pt]
$a_\mu$  & 0.001\,165\,92 &   Ref. \cite{codata22}  \\
$(1+a_\mu)\,\delta^{(1)}_\mathrm{evp}$ & 0.006\,082\,37 & Eq. (\ref{15}) \\
$(1+a_\mu)\,\delta^{(2)}_\mathrm{evp}$ & 0.000\,061\,32& Eq. (\ref{18}) \\
$\delta^{(2)}$ &  -0.000\,016\,34   & Eq. (\ref{90}), Ref.\cite{eides:01} \\
$\delta^{(3)}$ &  -0.000\,007\,10   & Eq. (\ref{91}), Ref.\cite{eides:01, heplus} \\
$\delta^{(3)}_\mathrm{rel,evp}$ & 0.000\,001\,15 & Eq. (\ref{33}) \\
$\delta^{(3)}_\mathrm{\mu vp,evp}$ & 0.000\,001\,17  & Eq. (\ref{86}) \\
$\delta^{(3)}_\mathrm{se,evp1}$ & -0.000\,001\,70   & Eq. (\ref{88}) \\
$\delta^{(3)}_\mathrm{se,evp2}$ & -0.000\,000\,70  & Eq. (\ref{Lee}), Ref.\cite{Lee_2023} \\
$\delta^{(1)}_\mathrm{fns}$ & -0.008\,237(21) & Eqs. (\ref{rZ1},\ref{rZ2}), Ref.\cite{Lin2022}\\
$\delta^{(1)}_\mathrm{rec}$  &   0.001\,672(3) & Eq. (\ref{94}), Ref.\cite{AA2022a} \\
$\delta^{(1)}_\mathrm{pol}$  &   0.000\,200\,6(52\,4) & Eq. (\ref{95}), Ref.\cite{Ruth2024} \\
$\frac{\alpha}{\pi}\,c_1\,\delta^{(1)}_\mathrm{fns}$ & -0.000\,033\,12& Eq. (\ref{96})\\
$\frac{\alpha}{\pi}\,c_1\,\delta^{(1)}_\mathrm{rec}$ & 0.000\,006\,72& Eq. (\ref{96}) \\
$\delta^{(2)}_\mathrm{evp,fns}$ & -0.000\,149\,81& Eq. (\ref{65})\\
$\delta^{(2)}_\mathrm{evp,rec}$ & 0.000\,025\,24 &  Eq. (\ref{66})\\ 
$\delta^{(2)}_\mathrm{\mu vp,fns}$ & -0.000\,025\,10 & Eq. (\ref{70})\\
$\delta^{(2)}_\mathrm{\mu vp,rec}$ & -0.000\,001\,18 & Eq. (\ref{71}) \\ 
$\delta_\mathrm{se,fns}^{(2)}$  & 0.000\,016\,18 & Eq. (\ref{80}) \\
$\delta_\mathrm{se,rec}^{(2)}$  &0.000\,016\,48 & Eq. (\ref{83})  \\
$\delta_\mathrm{rel,fns}^{(2)}$  &-0.000\,050\,85  & Eq. (\ref{101}), Ref.\cite{muD} \\
$\delta_\mathrm{rel,rec}^{(2)}$  & 0.000\,118\,86  &  Eq. (\ref{recoil}), Ref.\cite{Hevler:26}  \\
$\delta_\mathrm{rel,rec2}^{(2)}$  & 0.000\,000(12)  &  $ (Z\,\alpha)^2\,(m/M)^2$ \\
$\delta_\mathrm{rel,rec,fns}^{(2)}$  & 0.000\,000(12)  & $(Z\,\alpha)^2\,m^2/M\,r_Z$\\
$\delta^{(2)}_\mathrm{hvp}$ & 0.000\,011\,80 (8)& Eq. (\ref{hvp}), Ref.\cite{Lensky2026} \\
$\delta^{(3)}_\mathrm{fns}$ & 0.000\,000(2) & $\alpha^3\,m\,r_Z$ \\
$\delta_\mathrm{weak}$ &0.000\,011\,99 & Eq. (\ref{weak}), Ref.\cite{Eides96} \\ \\
$\delta$ & 0.000\,869\,(60)& Total value \\ 
$\frac{m_\mu}{m_e}\,[\delta_\mathrm{exp}(\mathrm{H}) - \delta(\mathrm{H})]$ &0.000\,133 & Sec. IX \\
$\delta_\mathrm{corr}$ & 0.001\,002\,(30)& Corrected total value \\ 
\end{tabular}
\end{ruledtabular}
\end{table}

The ground-state hyperfine splitting of $\mu$H  is conveniently represented as
\begin{align}
E_\mathrm{hfs} =&\ E_F\,(1+\delta)\,, \label{89}
\end{align}
where the dimensionless $\delta$ is the sum of various contributions  listed in Table I.
Let us now explain the meaning of all $\delta$ contributions.
$\delta^{(1)}_\mathrm{evp}$ in Eq. (\ref{15}) and $\delta^{(2)}_\mathrm{evp}$ in Eq. (\ref{18}) 
are one- and two-loop EVP corrections to the contact Fermi (spin-spin) interaction. 
$\delta^{(2)}$ and $\delta^{(3)}$ are QED corrections, which are exactly the same as in the electronic case \cite{eides:01},
\begin{align}
\delta^{(2)}&\  =\frac{3}{2}\,(Z\,\alpha)^2 + \alpha\,(Z\,\alpha) \Bigl(\ln(2)-\frac{5}{2}\Bigr)\,, \label{90} \\
\delta^{(3)}&\  = \frac{\alpha\,(Z\,\alpha)^2}{\pi}\!
\Big[\!-\frac{8}{3}\ln(Z\,\alpha)\Bigl(\ln(Z\,\alpha)-\ln(4)+\frac{281}{480}\Bigr)
\nonumber \\ &\
+ 17.122\,338\,751\,3-\frac{8}{15}\,\ln(2)+\frac{34}{225}\Bigr]
\nonumber \\ &\
+ \frac{\alpha^2\,(Z\,\alpha)}{\pi}\,0.770\,99(2) \,.  \label{91}
\end{align}
$\delta^{(3)}_\mathrm{rel,evp}$ in Eq. (\ref{33}) is the relativistic correction to the  one-loop EVP.
$\delta^{(3)}_\mathrm{\mu vp,evp}$ in Eq. (\ref{86}) is the combined $\mu$VP and EVP correction.
$\delta^{(3)}_\mathrm{se,evp1}$ in Eq. (\ref{88}) is the combined SE and EVP correction to the Coulomb interaction.
$\delta^{(3)}_\mathrm{se,evp2}$ is the EVP correction on the SE photon.
It was calculated by Krachkov and Lee in Ref. \cite{Lee_2023} with the following result:
\begin{align}
\delta^{(3)}_\mathrm{se,evp2} =&\ \frac{\alpha^2\,(Z\,\alpha)}{\pi}\,\biggl[
\biggl(-\frac{13}{6}+\frac{2}{3}\,\ln 2\biggr)\,\ln\frac{m_\mu}{m_e} 
\nonumber \\ &\
+ \frac{379}{72}-\frac{14}{9}\,\ln 2 -\frac{\pi^2}{18} +\frac{1}{3}\,\ln^2 2 
\nonumber \\ &\
-\frac{72}{25}\, \Gamma\bigg(\frac{3}{4}\bigg)^2\,\Gamma\bigg(\frac{5}{4}\bigg)^{-2}\sqrt{\frac{m_e}{m_\mu}} 
+O\bigg(\frac{m_e}{m_\mu}\bigg)\biggr]
\nonumber \\ =&\
-0.70\;\mathrm{ppm}\,. \label{Lee}
\end{align}
$\delta^{(1)}_\mathrm{fns}$ is the finite nuclear size correction related to the Zemach radius:
\begin{align}
\delta^{(1)}_\mathrm{fns} =&\ -2\,Z\,\alpha\,m\,r_Z \label{rZ1},\\
r_Z =&\ 1.054(3), \label{rZ2}
\end{align}
where the numerical value is taken from Ref. \cite{Lin2022}.
$\delta^{(1)}_\mathrm{rec}$ is the leading nuclear recoil correction
\begin{align}
\delta^{(1)}_\mathrm{rec} =&\ 837.6^{+2.8}_{-1.0}\,\mathrm{ppm}  +2\,Z\,\alpha\,m\,r_Z\,\frac{m}{m+M}, \label{94}
\end{align}
where the numerical value is from Ref. \cite{AA2022a}.
Because $r_\mathrm{Z}$ is scaled here by the muon mass rather than the reduced mass, our recoil correction includes an additional term.
$\delta^{(1)}_\mathrm{pol}$ is the leading nuclear polarizability correction taken from Ref. \cite{Ruth2024},
\begin{align}
\delta^{(1)}_\mathrm{pol} =&\ 200.6(52.4)\,\mathrm{ppm}\,. \label{95}
\end{align}
It is convenient to separately consider  EVP corrections
to the square of the wave function at the origin $\phi^2(0)$
\begin{align}
\phi^2(0)_\mathrm{evp} =&\ \phi^2(0)\,\biggl( 1+ \frac{\alpha}{\pi}\,c_1 + \frac{\alpha^2}{\pi^2}\,c_2\biggr) \label{96}, \\
c_1 =&\ 1.73115 \label{97}\,, \\
c_2 =&\ 7.2558 \label{98}\,,
\end{align}
where $c_1$ and $c_2$ are one- and two-loop EVP correction respectively, see Ref. \cite{sgk:09}.
Consequently,  $\delta^{(1)}_\mathrm{fns}$ and $\delta^{(1)}_\mathrm{rec}$ receive corrections due to $\phi^2(0)_\mathrm{evp}$,
along with additional EVP corrections to the hard two-photon exchange, denoted by $\delta^{(2)}_\mathrm{evp,fns}$ in Eq. (\ref{65}) and $\delta^{(2)}_\mathrm{evp,rec}$ in Eq. (\ref{66}), respectively.  
$\delta^{(2)}_\mathrm{\mu vp,fns}$ is the $\mu$VP combined with FNS [see Eq. (\ref{70})].
$\delta^{(2)}_\mathrm{\mu vp,rec}$ is the $\mu$VP combined with REC [see Eq. (\ref{71})].
$\delta^{(2)}_\mathrm{se,fns}$ is the $\mu$SE combined with FNS [see Eq. (\ref{80})].
$\delta^{(2)}_\mathrm{se,rec}$ is the $\mu$SE combined with REC [see Eq. (\ref{83})].
$\delta^{(2)}_\mathrm{rel,fns}$ is the nonrecoil relativistic correction with FNS, given by \cite{muD} 
\begin{align}
\delta^{(2)}_\mathrm{rel,fns} =& \frac{4}{3}\,(m r_p Z \alpha)^2\bigg[
\gamma -1 +\ln(2\, mr_{pp}\,Z \alpha) + \frac{1}{4}\,\frac{r_{m}^2}{r_p^2} \bigg]\,.\label{relfns}
\end{align}
For the dipole parametrization of the nuclear form factors
 \begin{align}
 	r_{m} =r_p\,,\ \
 	r_{pp}/r_p = 5.274\,565\ldots\,,
 \end{align}
 where $r_p = 0.840\,60\, (39)$ fm is the proton charge radius; therefore,
 \begin{align}
\delta^{(2)}_\mathrm{rel,fns}(\mathrm{\mu H}) =& -50.85 \text{ ppm} \label{101}
\end{align}
$\delta^{(2)}_\mathrm{rel,rec}$ is the relativistic recoil correction  \cite{Hevler:26} 
\begin{align}
&\delta_{\text{rel,rec}}^{(2)} = \frac{m}{M} \frac{(Z \alpha)^2}{1 + \kappa} \bigg[ \frac{65}{18} + \frac{13}{18} \kappa + \frac{31}{36} \kappa^2  \label{recoil}
\\ \notag
&\, \,- \left( 8 + 2\kappa - \frac{1}{4} \kappa^2 \right) \ln 2 - \left( 2 + 2\kappa + \frac{7}{4} \kappa^2 \right) \ln (Z \alpha) \bigg].
\end{align}
$\delta^{(2)}_\mathrm{rel,rec2}$ is the relativistic second-order recoil correction $\sim (m/M)^2$ for which we provide only an estimate.
$\delta^{(2)}_\mathrm{rel,rec,fns}$ is the relativistic recoil finite nuclear size correction and is also estimated only.
$\delta_\mathrm{hvp}$ represents the hadronic VP, and we adopt the result from  Ref. \cite{Lensky2026}
\begin{align}
\delta_\mathrm{hvp} =&\ 11.80(8)\;\mathrm{ppm} \,. \label{hvp}
\end{align}
$\delta^{(3)}_\mathrm{fns}$ is FNS correction at the order $\alpha^3$, for which we provide estimation only.
Finally, $\delta_\mathrm{weak}$ is the nonrecoil weak interaction correction for a point nucleus \cite{Eides96}
\begin{align}
\delta_\mathrm{weak}(\mu\mathrm{H}) =&\ \frac{m_\mu}{m_e}\,58\cdot 10^{-9}\,. \label{weak}
\end{align}

\section{$\mu$H vs H}
One can use HFS measurement in H to extract  $\delta^{(1)}_\mathrm{nuc}(\mathrm{H}) 
= \delta^{(1)}_\mathrm{fns}(\mathrm{H}) + \delta^{(1)}_\mathrm{rec}(\mathrm{H}) + \delta^{(1)}_\mathrm{pol}(\mathrm{H})$ 
and use it to improve theoretical predictions for $\mu$H.
This is possible because all other contributions to the HFS in H are well known.
Therefore, let us consider the proton structure corrections to the scaled difference between $\mu$H and H
\begin{align}
\Delta =&\ \delta(\mu\mathrm{H}) -\frac{m_\mu}{m_e}\,\delta(\mathrm{H})\,.
\end{align}
One can expect that the nuclear structure contributions, as well as their associated uncertainties, cancel out to a high degree in this difference. 
The nuclear part is
\begin{align}
\Delta_\mathrm{nuc} =&\ \delta^{(1)}_\mathrm{nuc}(\mu\mathrm{H}) - \frac{m_\mu}{m_e}\,\delta^{(1)}_\mathrm{nuc}(\mathrm{H})
\nonumber \\ =&\ \Delta_\mathrm{fns} +  \Delta_\mathrm{rec} +  \Delta_\mathrm{pol} \label{Delta1}\,. 
\end{align}
The FNS contribution $\Delta_\mathrm{fns}$ vanishes exactly; so does the related uncertainty. The recoil contribution
$\Delta_\mathrm{rec}$ 
\begin{align}
\Delta_\mathrm{rec} =&\ 
 \delta^{(1)}_\mathrm{rec}(\mu\mathrm{H}) - \frac{m_\mu}{m_e}\,\delta^{(1)}_\mathrm{rec}(\mathrm{H})
 \nonumber \\ =&\  0.000\,578(1),
\end{align}
is decreased by a factor of 3 compared to $\delta^{(1)}_\mathrm{rec}(\mu\mathrm{H})$. Thus, 
we assume that its uncertainty is also decreased by this factor, rendering it negligible.
The proton polarizability contribution
\begin{align} 
\Delta_\mathrm{pol} =&\ 
 \delta^{(1)}_\mathrm{pol}(\mu\mathrm{H}) - \frac{m_\mu}{m_e}\,\delta^{(1)}_\mathrm{pol}(\mathrm{H})
\nonumber \\ =&\  -0.000\,025(25),
\end{align}
is decreased by a factor of 8 compared to $\delta^{(1)}_\mathrm{pol}(\mu\mathrm{H})$, and we estimate its uncertainty to be as large as the entire value of $\Delta_\mathrm{pol}$.
It is therefore  less than half the size of the uncertainty in $\delta_\mathrm{pol}(\mu\mathrm{H})$. We believe that this can be further improved with a more detailed analysis.

Let us  now  consider the theoretical result for $\delta(\mu\mathrm{H})$ and replace it with
\begin{align}
\delta_\mathrm{corr}(\mu\mathrm{H}) =&\ \delta(\mu\mathrm{H}) + \frac{m_\mu}{m_e}\,[\delta_\mathrm{exp}(\mathrm{H}) - \delta(\mathrm{H})]
\nonumber \\ =&\
\Delta +  \frac{m_\mu}{m_e}\,\delta_\mathrm{exp}(\mathrm{H}) 
\nonumber \\ =&\ 0.001\,002(30)\,,
\end{align}
where the correction without uncertainty is \cite{Hevler:26}
\begin{align}
\frac{m_\mu}{m_e}\,[\delta_\mathrm{exp}(\mathrm{H}) - \delta_\mathrm{theo}(\mathrm{H})] =&\ 0.000\,133\label{Delta2}\,.
\end{align}
 The uncertainty of $\delta_\mathrm{corr}(\mu\mathrm{H})$ is obtained from the uncertainty of
$\Delta$, which originates from the 25 ppm of $\Delta_\mathrm{pol}$, the 1 ppm of  $\Delta_\mathrm{rec}$,
and all  other uncertainties  listed in Table I.
The corrected theoretical predictions for $\mu$H is therefore
\begin{align}
E_\mathrm{hfs}(\mu\mathrm{H}) =&\ E_F\,(1+\delta_\mathrm{corr}) = 0.182\,626(5)\;\mathrm{eV}. \label{111}
\end{align}
It is interesting to note that this value is close to the one obtained from the  nonrelativistic formula in Eq. (\ref{04})
with the muon magnetic moment  anomaly 
\begin{align}
E_F\,(1+a_\mu) =&\ 0.182\,656\;\mathrm{eV},
\end{align}
indicating a tendency for higher-order corrections to cancel out.

In comparison to the previous work on this topic by Faustov and Martynenko in Ref. \cite{Martynenko2004},
we observe an agreement only for the one-loop EVP correction.  Moreover, it was difficult to perform further comparison  term by term, 
due to inconsistent classification of their corrections, but whenever we were able to compare, we observed a disagreement.
In contrast, our result is in good agreement with the value $E_\mathrm{hfs}(\mu\mathrm{H})  = 0.182\,636(8)$ eV,  presented by Antognini {\em et al.}
 in Ref. \cite{AA2022b}, but we were also not able to perform comparison term by term, due to lack of detailed results in that work. 

\section{Conclusions}
We have accounted for all QED and recoil corrections larger than 1 ppm to the ground-state HFS of $\mu$H
and obtained numerical values or estimates, as summarized in Table I. Our result for $E_\mathrm{hfs}(\mu\mathrm{H})$ in Eq. (\ref{111}) 
corresponding to $\lambda = 6.788\,97(19)\;\mu$m might serve
as a starting point to search for this hyperfine transition, which so far has not been observed.
The most important outcome, however, was the identification of corrections that still need to be calculated or improved in order to reach 1 ppm accuracy.
The first of these is the proton structure correction, and more precisely the weighted difference $\Delta_\mathrm{nuc}$ in Eq. (\ref{Delta1}),
which can be obtained much more accurately than $\delta^{(1)}_\mathrm{nuc}$ itself.
The second important contribution left for future work is the $(Z\,\alpha)^2$ correction, evaluated without expansion in the mass ratio and
including the finite nuclear size. At present, we know the point nucleus values in the nonrecoil limit Eq. (\ref{90}), the leading recoil correction Eq. (\ref{recoil}),
and the FNS in the nonrecoil limit Eq. (\ref{101}). The omitted terms constitute the largest uncertainty besides $\Delta_\mathrm{nuc}$.
Their calculation is certainly feasible, but most importantly, the contents of Table I with various corrections should be independently verified. 

Our current theoretical predictions with incorporation of H HFS are presented in Eq. (\ref{111}).
It is clear that the HFS measurement in $\mu$H with 1 ppm accuracy will stand as a highly significant test
of fundamental interactions when combined with H HFS or alternatively will serve for the accurate determination of the proton Zemach radius.

\acknowledgments
We acknowledge inspiration by Aldo Antognini and wish to thank Vadim Lensky for the $\delta_\mathrm{hvp}$ value.

\appendix

\section{Lepton self-energy integrals}
The lepton factors  $t_\mathrm{1se}$ and $t_\mathrm{2se}$ can be expressed in terms of scalar integrals in the form
\begin{align}
    f(n,m,l) = \int \frac{\ddD{k}}{i\pi^{D/2}} \frac{1}{[(k - t)^2]^n [k^2-1]^m [(k + q)^2-1]^l},
\end{align}
where $t^2 = 1$ and $D=4-2\,\epsilon$.
Using integration by parts identities \cite{remiddi1, remiddi2}, one can algebraically reduce all powers $n,m$, and $l$ to 0 and 1.
These integrals with lowest powers can all be performed using the Feynman parameters approach. 
Neglecting $\mathcal{O}(\epsilon)$, they are
\begin{align}
    f(n,0,0) =&\ 0\,,\\
    f(0,1,0)=&\ \frac{1}{\epsilon}+1-\gamma_{E},\\
    f(0,2,0)=&\ \frac{1}{\epsilon}-\gamma_{E}, \\
    f(0,0,1) =&\ \frac{1}{\epsilon}+1-\gamma_{E},\\
    f(0,0,2) =&\ \frac{1}{\epsilon}-\gamma_{E}, \\
    f(1,1,0)=&\  \frac{1}{\epsilon}+2-\gamma_{E},\\
    f(1,2,0) =&\ \frac{1}{2\,\epsilon} - \frac{\gamma_{E}}{2}, \\
    f(0,1,1)=&\  \frac{1}{\epsilon}-\gamma_{E}+2-2\,\sqrt{\frac{4}{q^{2}}-1}\,\arcsin\left(\frac{q}{2}\right), \\
    f(1,0,1)=&\ \frac{1}{\epsilon}-\gamma_{E}+2+\frac{1-p^2}{p^2}\ln(1-p^2),\\
    f(0,2,1) =&\ -\frac{2}{q^2}\,\bigg(\frac{4}{q^2} - 1\bigg)^{-1/2}\,\arcsin\left(\frac{q}{2}\right), \\
    f(1,2,1) =&\ \frac{1}{p^2-1}\,\bigg[\frac{1}{2\,\epsilon} - \frac{\gamma_{E}}{2} - \ln(1 - p^2)
    \nonumber \\ &\
    + 2\,\bigg(\frac{2}{q^2} - 1\bigg)\,\bigg(\frac{4}{q^2} - 1\bigg)^{-1/2}\!\arcsin\left(\frac{q}{2}\right) \bigg],\\
    f(1,1,1) =&\ -J(-q^2,q^0),
\end{align}
where $\gamma_E$ is the Euler constant, $p = t+q$, and $q^0 = q\,t$. The master integral $J$ is 
\begin{align}
J(-q^2,q^0) =&\ -\int \frac{d^4 k}{\pi^2\,i}\;\frac{1}{k^2}\,\frac{1}{(t-k)^2-1}\;\frac{1}{(p-k)^2-1}
\nonumber \\ =&\
 \int_0^1 du \;\frac{1}{1-u(1-u)\,q^2 - u(1-p^2)}
 \nonumber \\ &\times
 \ln\left(\frac{1-u(1-u)q^2}{u(1-p^2)}\right). \label{B01}
\end{align}
The particular form at $q^0=0$ of the master integral $J(q^2) \equiv J(q^2,0)$ is
\begin{align}
J(q^2) =&\ \int_0^1 du \;\frac{1}{1-u^2\,q^2}\;\ln\left(\frac{1+u(1-u)q^2}{u\,q^2}\right) \nonumber 
\\
=&\  1 + \frac{5\,q^2}{18} + \frac{11\,q^4}{150}  
 -\bigg(1 + \frac{q^2}{3} + \frac{q^4}{5}\bigg)\,\ln(q^2) + \ldots \nonumber 
\\
=&\ 
\frac{2}{q^2}+ \frac{2}{9\,q^4}  
 +\bigg(\frac{1}{q^2}  - \frac{5}{3\,q^4}  \bigg)\,\ln(q^2) + \ldots
\end{align}
The derivative of $J(q^2)$ satisfies
\begin{align}
\frac{\partial}{\partial q}  \big[ q J(q^2)\big] =&\ -\frac{4}{q^2}\,\frac{\mathrm{arcsinh}\frac{q}{2}}{\sqrt{1+\frac{4}{q^2}}} - \frac{\ln q^2}{1 - q^2}\,.
\end{align}
Other properties of $J$ can be found in Appendix B of Ref. \cite{pachucki:25}

\end{document}